\documentclass[12pt,letterpaper,oneside,onecolumn]{article}
\usepackage[]{fontenc}
\usepackage[dvips]{epsfig}
\usepackage[]{rotating}
\newtheorem{sfig}{Figure}[section]

\begin{document}

\begin{center}
\large
{\bf  Light $q\bar{q}$ Resonances: Spectrum and Decays}
\end{center}

\normalsize
\begin{center}
Stephen Godfrey \\
\small
{\em Ottawa-Carleton Institute for Physics, \\
Department of Physics, Carleton University, Ottawa} Canada K1S 5B6 \\
\end{center}

\begin{center}
\large
{\bf Abstract }
\normalsize
\end{center}
\begin{quotation}
\begin{quotation}
I give a brief introduction to light $q\bar{q}$ resonances. 
The properties of $q\bar{q}$ states are discussed in the context 
of the constituent 
quark model, including predictions for mass and partial decay widths.
Problems and puzzles in the meson spectrum are pointed out with an 
emphasis on how an 8+~GeV photon beam can address these.  I also point 
out some areas where I believe theoretical study would be useful.
\end{quotation}
\end{quotation}

\section{Introduction}

Meson physics has been an important subject in subatomic physics for 
many years.  It remains a relevant and important subject for many 
reasons.  To begin with, despite the fact that mesons have been studied 
for many years, one would be hard pressed to say that they are well 
understood:  Too many of the $q\bar{q}$ states predicted by the quark 
model have yet to be seen -- only the ground state S and P wave 
multiplets are filled;  there are puzzles in many $J^{PC}$ sectors 
with more states than expected by the quark model; and there are no 
confirmed quark model exotic states.  

There are many reasons to study conventional $q\bar{q}$ mesons.  
As we just said they are still relatively poorly understood.  Until 
more states have been sited and their properties measured we cannot 
say that the subject is complete.  In addition, there is a large 
effort devoted to identifying non-$q\bar{q}$ states.  Many of 
the strongest candidates have quantum numbers consistent with 
$q\bar{q}$ expectations.  Thus, to unambiguously identify a candidate 
as a non-$q\bar{q}$ resonance requires a good understanding of 
conventional meson properties.  A better understanding of mesons will 
eventually lead to a better understanding of Quantum Chromodynamics, 
the theory of the strong interactions.  Because Quantum Chromodynamics 
is the only strongly interacting field theory that we currently know 
of and that we can compare theoretical predictions to experimental 
measurements, improving our knowledge of QCD will lead to a better 
understanding of strongly interacting field theories.  This 
could have important implications elsewhere.  For example,  one 
possiblility of electroweak symmetry breaking is that at high energies 
the weak interactions become strong.  

Although the quark model very successfully describes much of hadron 
spectroscopy \cite{gi,godfrey85,sg,ci} 
we know that it only includes the lowest order components of  the Fock 
space expansion.  One way of including higher Fock space components in 
the context of the quark model is 
to include coupling to decay channels in a coupled channel approach
\cite{jw} 
while another parametrizes these effects as potentials in final state 
interactions \cite{b-s}.
Recent work finds that including higher components 
of the Fock space can lead to large changes of the hadron properties
\cite{i-c}.
For example the work of Weinstein \cite{jw} 
has shown that coupling the $q\bar{q}$ states to mesons can lead to 
shifts in the resonance pole.  Likewise, Geiger and Swanson 
\cite{g-s} have found that including final state interactions in 
hadron decays can have an effect on decay properties.  It is becoming 
clear that these effects must be included as part of our understanding 
of mesons spectroscopy.

Finally we note that to unravel the meson spectrum it is necessary to 
study as many channels as possible.  Only by comparing the observed 
couplings to those  predicted for specific states in many channels can 
we come to a meaningful conclusion about the nature of an observed 
resonance.  For example, in addition to
couplings to hadronic states, the $2\gamma$ 
couplings contribute to our knowledge of observed states.  By putting 
all these bits of information together
we can eventually come to a conclusion 
about the nature of observed resonances.

\section{Meson Spectroscopy}

We will use the the constituent quark model as a template for 
conventional $q\bar{q}$ states as it consistently gives a good 
description of hadron properties \cite{gi}.  
Our ultimate goal is to find evidence 
for non-$q\bar{q}$ states.  Providing a good template is an important 
starting point against which we can compare the properties of exotic 
candidates.  Discrepanices between the predictions of the quark model 
and observed resonances suggest areas where new physics effects may 
play a role.  Thus we need a good understanding of the quark model and 
in particular we need a much better understanding of light meson 
spectroscopy than currently exists.  For example, there are numerous 
missing $q\bar{q}$ states:  Only the 1S and 1P multiplets
are filled and  beyond 
these, the radially and orbitally excited multiplets are 
woefully incomplete.  In addition, there are numerous puzzles that must 
be resolved such as more scalar and tensor mesons
than can be explained as $q\bar{q}$ resonances.
Thus,  we need to find more of the missing $q\bar{q}$ states and 
resolve the various puzzles before we can say we have an adequate 
understanding of light meson spectroscopy. 

\subsection{Meson Quantum Numbers}

In the constituent quark model the quark and antiquark spins are 
combined to give total spin $S$ which is then combined with orbital 
angular momentum $L$ to give the total angular momentum $J$.  In 
spectroscopic notation the resulting state is denoted by 
$^{2S+1}L_J$ 
with S for L=0, P for L=1, D for L=2, and F, G, H, for L=3,4,5 etc.
Parity is given by $ P(q\bar{q},L) = (-1)^{L+1} $ and neutral 
self-conjugage mesons which
are also eigenstates of C-Parity which is given by
$ C(q\bar{q},L,S)= (-1)^{L+S} $.
The resulting quantum numbers for the low lying mesons are given in 
Table I.

\begin{table}[h!]
\caption{$q\bar{q}$ composites for $L\leq 2$}
\begin{center}
\begin{tabular}{cclc} \hline
L & S & $J^{PC}$ & Notation \\ \hline
0 & 0 & $0^{-+}$ & $^1S_0$ \\
  & 1 & $1^{--}$ & $^3S_1$ \\
1 & 0 & $1^{+-}$ & $^1P_1$ \\
  & 1 & $0^{++}$ & $^3P_0$ \\
  &   & $1^{++}$ & $^3P_1$ \\
  &   & $2^{++}$ & $^3P_3$ \\
2 & 0 & $2^{-+}$ & $^1D_2$ \\
  & 1 & $1^{--}$ & $^3D_1$ \\
  &   & $2^{--}$ & $^3D_2$ \\
  &   & $3^{--}$ & $^3D_3$ \\
\hline
\end{tabular}
\end{center}
\end{table}

Not all combinations of $J^{PC}$ can be made out of $q\bar{q}$.  For 
example the quantum numbers $0^{--}$ are forbidden as is the whole 
sequence $0^{+-}$, $1^{-+}$, $2^{+-} \ldots$.  These non-conventional 
quantum numbers are good signatures for non-quark model states. 

\subsection{The Confinement Potential}

In the quark model a phenomenological potential is used in a 
Schrodinger-type equation which is solved to give the meson masses.  
The potential is typically of the form:
\begin{equation}
V(r) = - \frac{4}{3} \frac{\alpha_s(r)}{r} + br.
\end{equation} 
The parameters of the potential 
are the strong Coulomb strength, $\alpha_s$, and the slope of the 
linear confining piece, $b$. Historically this form was motivated by 
the expectations of one-gluon-exchange at short distance and a linear 
confining potential at large separation which were fitted by 
phenomenological fits to the data.  This form has been verified by 
lattice calculations in the heavy quark limit and using lattice 
derived potentials gives reasonably good results for the $b\bar{b}$ and
$c\bar{c}$ spectra \cite{lattice}. 
Although the constituent quark model using this 
potential gives a good description of hadrons with light quarks there 
is no rigorous theoretical justification for why it should work so 
well in the light quark sector.

In Fig. 1 the confining potential is plotted along with the rms radii 
for various mesons.  The heavy quarkonium have smaller radii and are 
therefore more sensitive to the short range colour-Colomb interaction 
while the light quarks, especially the orbitally excited mesons, have 
larger radii and are more sensitive to the long range confining 
interaction.  Thus, measurement of both heavy quarkonium and mesons 
with light quark content probe different regions of the confinement 
potential and complement each other.  

\begin{figure}[h!]
\begin{center}
\centerline{\epsfig{file=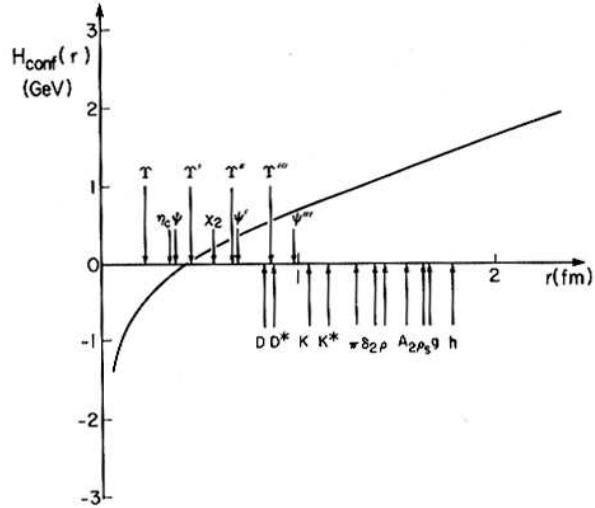,width=8cm,clip=}}
\end{center}
\caption{The $q\bar{q}$ potential in a colour singlet meson.  Also 
shown is the rms $q\bar{q}$ separation in some representative mesons.  
From Ref. \cite{gi}.}
\end{figure}

Another way of showing this is 
given in Fig. 2 which plots the regge trajectories of the isovector, 
strange, and strangeonium mesons on a Chew-Frautschi plot 
\cite{godfrey85}. 
 The 
masses of the observed high $L$ states fall on these trajectories.
The straight line behavior of this 
plot reflects the linear nature of the confining potential.  This can 
be seen by examining the relativistic Schrodinger-like equation:
\begin{equation}
[\sqrt{p^2 + m_q^2} + \sqrt{p^2 + m_{\bar{q}}^2} - \frac{4}{3} 
\frac{\alpha_s}{r} +br + C] |\psi \rangle = E |\psi \rangle .
\end{equation}
In the large-$L$ limit this becomes
\begin{equation}
(L/r + br) \psi(r) = E \psi(r)
\end{equation}
which has a classical minimum at $r= (L/b)^{1/2}$.  Upon substitution 
this yields $E^2 = 4 b L$, the observed behavior.

\begin{figure}[h!]
\begin{center}
\centerline{\epsfig{file=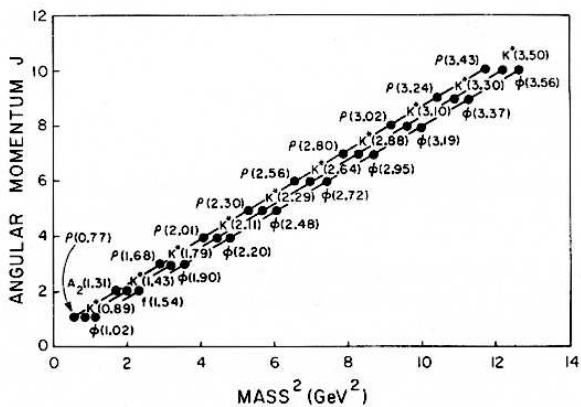,width=8cm,clip=}}
\end{center}
\caption{Chew-Frautschi plot for the $\rho$, $K^*$, and $\phi$ trajectories. 
 From ref. \cite{godfrey85}.}
\end{figure}

\subsection{The Spin Dependent Potentials}

In addition to the confining potential there are spin dependent 
pieces of the Hamiltonian.  The first piece that we discuss is the 
hyperfine interaction which includes a contact piece and a tensor 
piece.  It arises from colour-magnetic interactions originating in the 
short range one-gluon-exchange in analogy to similar effects in atomic 
physics originating in QED.  The hyperfine Hamiltonian 
is given by the following equation:
\begin{equation}
H_{ij}^{hyp} = \frac{4\alpha_s(r)}{3m_i m_j} \left\{ \frac{8\pi}{3}
\vec{S}_i \cdot \vec{S}_j \delta^3 (r_{ij}) + \frac{1}{r_{ij}^3} \left[
\frac{3\vec{S}_i \cdot \vec{r}_{ij}\vec{S}_j \cdot
\vec{r}_{ij}}{r_{ij}^3} - \vec{S}_i \cdot \vec{S}_j \right] \right\} 
\end{equation}
It is responsible for the $^3S_1 - ^1S_0$ splittings; for example 
the $\rho-\pi$, $K^* -K$, $D^*-D$, and $J/\psi -\eta_c$ splittings.  
The short range nature of the contact interaction is supported by the 
small splittings between the centre of gravity of the $^3P_J$ and the 
$^1P_1$ multiplets.  For P-waves the wavefunction at the origin is 
zero so that the expectation value of the $\delta$-function in the 
contact term is zero.  If the $\vec{S}\cdot \vec{S}$ term were long 
range one would expect splittings between the $^3P_J$ and the 
$^1P_1$ multiplets of the same order as the $^3S_1 - ^1S_0$ splittings.

The second spin-dependent contribution is the spin-orbit interaction.  
There are two contributions to the spin-orbit interaction.  The first 
is a purely relativistic effect.  An object with spin moving in a 
central potential will undergo Thomas precession.  This term is given 
by:
\begin{eqnarray}
H_{ij}^{S.O.(t.p.)} & = & -\frac{1}{2r_{ij}} 
\frac{\partial V(r)}{\partial r_{ij}} 
\left( \frac{\vec{S}_i}{m_i^2} +
\frac{\vec{S}_j}{m_j^2} \right) \cdot \vec{L} .
\end{eqnarray}
In addition there is a colour magnetic spin-orbit term arising from 
one-gluon-exchange.  This term is given by:
\begin{eqnarray}
H_{ij}^{S.O.(c.m.)} &=& \frac{4\alpha_s(r)}{3r_{ij}^3} \left(
\frac{1}{m_i} + \frac{1}{m_j} \right) \left( \frac{\vec{S}_i }{m_i} +
\frac{\vec{S}_j}{m_j}\right)\cdot \vec{L} .
\end{eqnarray}
The spin-orbit Hamiltonian contributes to splittings within $^3L_J$ 
multiplets for $L\neq 0$.  It also contributes to the mixing of 
$^3L_J -^1L_J$ states when $m_q \neq m_{\bar{q}}$.  For unequal mass
quark and antiquark $C$-conjugation is no longer a good quantum number 
and states with different $S$ but the same total $J$ can mix 
\cite{gk,bgp}.
Note that the two terms contribute with  opposite sign.  At short 
distance the colour magnetic contribution with $r^{-3}$ dominates 
while at large average separation the Thomas precession contribution 
dominates.  Thus, at high orbital (and radial) excitation where the 
$q-\bar{q}$ pair have on average larger separation, we expect the 
triplet multiplets to invert relative to the ordering of the low 
excitation multiplets \cite{godfrey}.  By this we mean that 
$M(^3L_{L+1}) > M(^3L_{L}) > M(^3L_{L-1}) $ at low $L$ while 
$M(^3L_{L-1}) > M(^3L_{L}) > M(^3L_{L+1}) $ at high $L$.  Because 
the details of this multiplet inversion depend on the confinement 
potential measuring the masses of excited mesons gives us information 
about the confinement potential that cannot be obtained elsewhere.

\subsection{The Strangeonium Spectrum}

\begin{figure}[t!]
\begin{center}
\epsfig{file=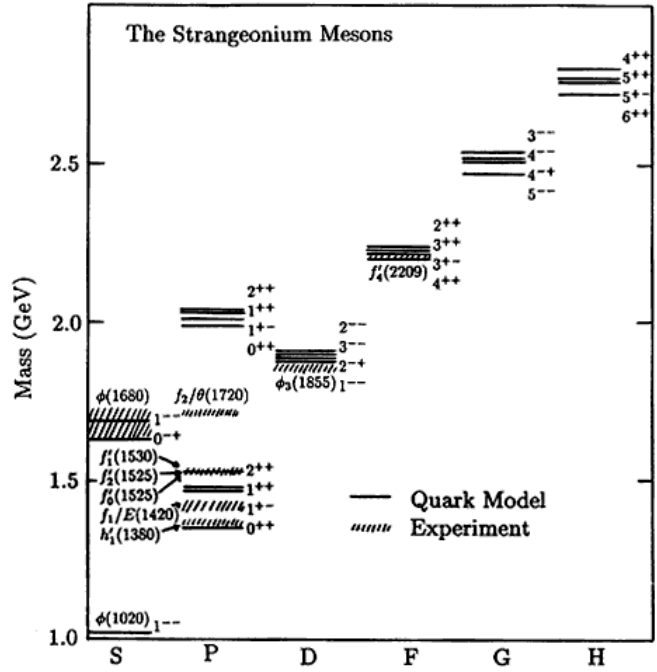,width=9.0cm,clip=}
\end{center}
\caption{The Strangeonium mass spectrum.  
The mass predictions come from Ref. \cite{gi}.}
\end{figure}

Most of the details we just discussed are summarized in the plot of 
the strangeonium mass spectrum shown in Fig. 3.
The first observation is the regularity of the spectrum.  There is 
good agreement for the masses of the leading orbital excitations which 
supports the linearity of the confining potential.  Extracting the 
$^3S_1 - ^1S_0$ splitting from the $\phi -\eta'$ splitting is 
non-trivial because of the large annihilation mixing between the 
$0^{-+}$ $ns$ and $s\bar{s}$ components \cite{isgur,gi}.  
Nevertheless comparing the 
$^3S_1 - ^1S_0$ to the $^3P_J - ^1P_1$ splittings is consistent with 
the expected short distance contact interaction arising from
one-gluon-exchange.  What is striking about this figure is how little 
information exists about higher orbitally excited and radially excited
multiplets.  Only the ground state 1S and 1P multiplets are complete 
and even in the 1P multiplet questions exist.  As has been 
pointed out already, the study of the properties of these mesons can 
yield a better understanding of the underlying theory.
As to the problems alluded to,  one 
notes that there are currently two candidates for the $^3P_1$ state, 
the $f_1/E(1420)$ and the $f_1'(1530)$.  There is an ongoing discussion 
in the literature about the true nature of these states \cite{e-iota}.  
In addition, 
one notes the existence of the $f_J/\theta(1720)$ \cite{f1720}.  
Although it is not 
considered to be a candidate for the 1P $s\bar{s}$ multiplet it does 
not fit into the predicted $s\bar{s}$ spectrum.  Speculation exists in 
the literature as to whether this state may be a glueball \cite{f-glue} 
or perhaps a 
$V V$ molecule (vector meson-vector meson) \cite{vv}.   We can conclude that 
until much more is known about the $s\bar{s}$ spectrum we cannot say 
that we truly understand the $s\bar{s}$ mesons.

\subsection{Additional Comments on Quark Model Predictions}

In the previous sections I gave the conventional quark modeller's view 
about the meson spectrum.  Before proceeding I think it is important 
to introduce additional effects whose study is still in its infancy. 
The first of these effects is to include coupled channel effects in 
the study of mesons.  
In the simplest example Weinstein  \cite{jw} has shown that 
the multiplet degeneracy (ignoring spin dependent interactions)
can be broken by coupling the $q\bar{q}$ Hamiltonian to a meson-meson 
scattering Hamiltonian.  The coupling between the $q\bar{q}$ channel 
and the meson-meson channel is included via a $MM \leftrightarrow 
q\bar{q}$ transition operator.  The coupled channel Hamiltonian is 
given by:
\begin{equation}
\left( \begin{array}{cc} H_{AB}(r_{AB}) & V_{AB\leftrightarrow R}(r_{AB}) \\
V_{AB\leftrightarrow R}(r_{AB}) & H_S{R_{q\bar{q}}} (r_{q\bar{q}}) 
\end{array} \right ) 
\end{equation}
The different $J$ dependent angular momentum terms in $H_{MM}$ gives rise to 
shifts in the resonance pole resulting in multiplet splitting.
Weinstein's results for the strange P-wave mesons are given in Fig. 4.
The scattering curves for the $K_0$ and $K_2$ are obtained in $K\pi$ 
scattering while the $K_1$ curve is obtained from $K^*\pi$ 
scattering.  A complete analysis would of course include both spin 
dependent interactions in $H_{q\bar{q}}$ and all channels that could 
couple to the quantum numbers being studied.  Such a study may, for 
example, explain the strange axial meson mixing and the discrepancy 
between the quark model predictions and experimental measurements of 
the exicted $\rho$ meson properties.

\begin{figure}[h!]
\begin{center}
\epsfig{file=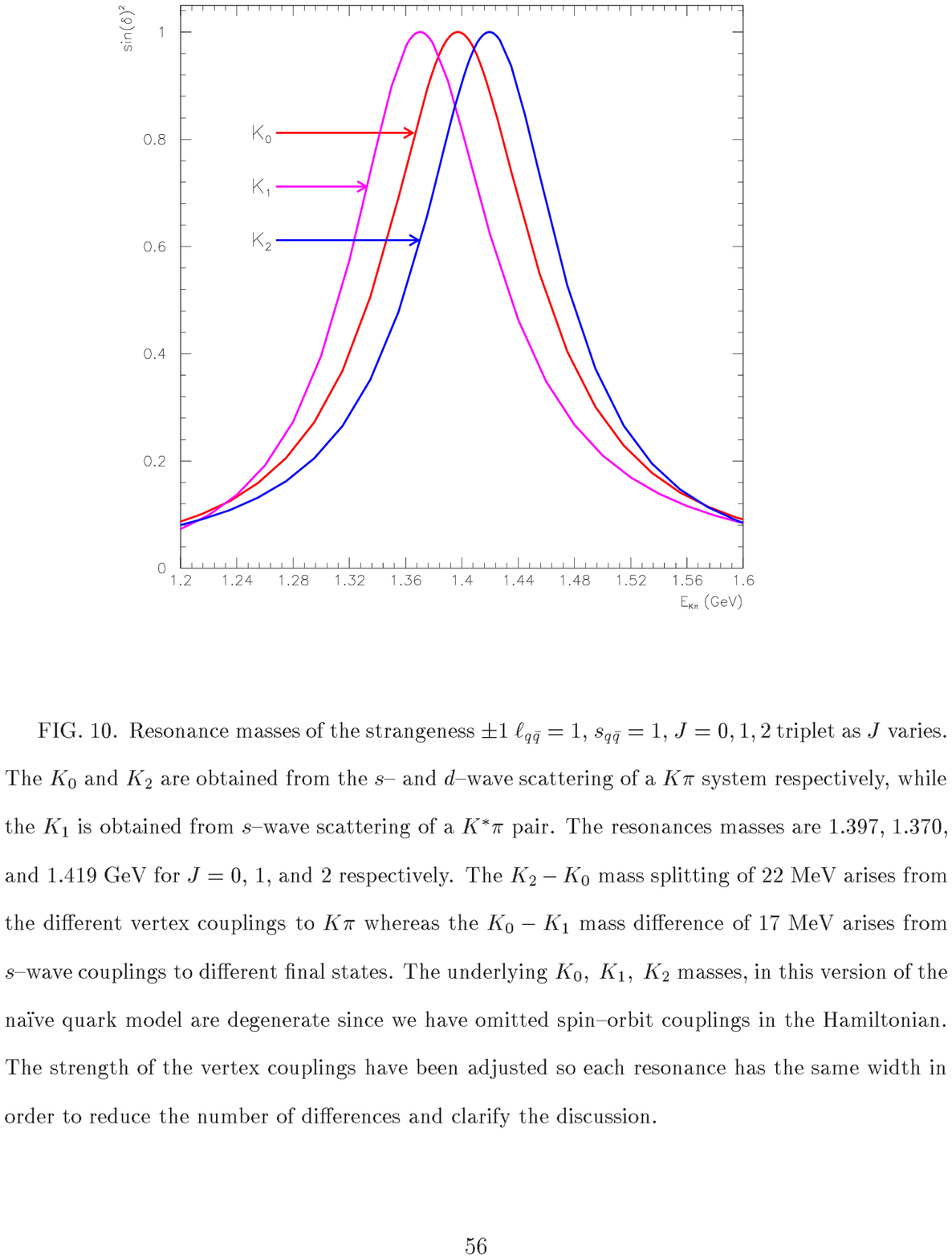,width=8.0cm,clip=}
\end{center}
\caption{Resonance masses of the strangeness $\pm 1$, 
$\ell{q\bar{q}}=1$, $S{q\bar{q}}=1$, $J=0, \; 1, \; 2$ triplet as $J$ 
varies.  The $K_0$ and $K_2$ are obtained from the $S-$ and $D-$wave 
scattering of $K\pi$ respectively while the $K_1$ is obtained from 
$s-$wave scattering of a $K^*\pi$ pair.  From Ref. \cite{jw}.}
\end{figure}

The second effect that might potentially make an important 
contribution to meson properties  is the contribution of final state 
interactions.  These may have an effect on the widths calculated in a 
hadron decay model but they should also be included in coupled channel 
calculations like those described above.  Ultimately one wants to 
perform a calculation that resembles as closely as possible what is 
measured by an experiment.  A systematic analysis of mesons which 
includes these contributions would clearly be a useful addition to the 
subject.

\subsection{Summary of Mesonic States}

Our knowledge of light meson spectroscopy is summarized in Fig. 5.  
Note that most radial and orbital excitations are 
missing.  This question must be addressed but it is likely due to a 
number of reasons.  In some cases the states are simply not produced 
in channels that have been easily accessable experimentally.  In other 
cases the states are too broad to be easily seen while in still other 
cases, 
although the states are not too broad, they decay via broad isobars 
making it difficult to reconstruct the resonance.  

\begin{figure}[h!]
\begin{center}
\epsfig{file=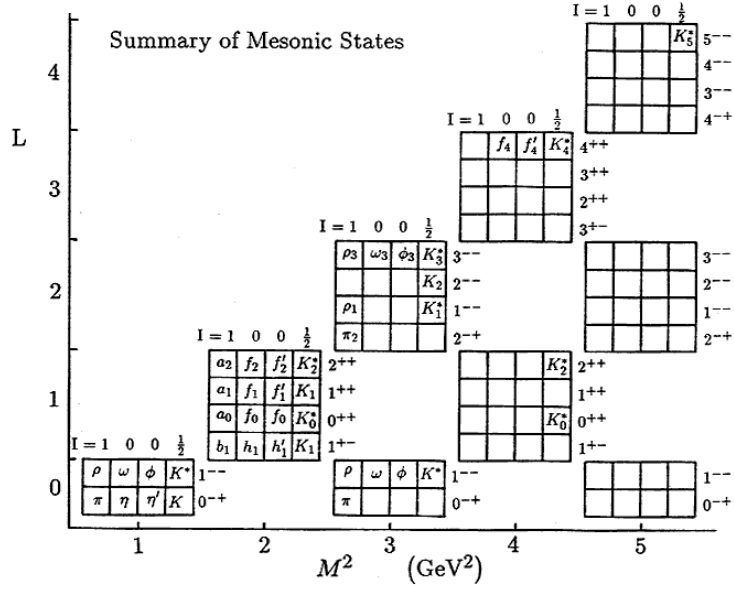,width=10.0cm,clip=}
\end{center}
\caption{Summary of mesonic states with light quarks.}
\end{figure}

In addition to the question of missing states there are numerous 
longstanding puzzles.  Some of these are:
\begin{description}
\item[$\eta/\iota(1440)$] The $\eta(1440)$ is seen 
in $J/\psi$ radiative decay and is observed to decay with a large 
partial width to $\gamma \rho$.  This implies that it is a mainly $ns$ 
radial excitation of the $\eta$.  However,  another candidate for the 
mainly $ns$ radial excitation is the $\eta(1295)$.  Recent 
measurements suggest that the $\eta(1440)$ may be split into two 
$0^{-+}$ states.
\item[$f_1/E(1440)$] This state was first reported in 1980.  It was 
regarded to be the $1^{++}$ $s\bar{s}$ state.  Since then there have 
been a number of observations of an axial meson with mass around 
1510~Mev.  The higher mass is more consistent with 
the $s\bar{s}$ P-wave multiplet than the $f_1(1440)$ mass.  As a 
consequence there is speculation that the $f_1(1440)$ is a hybrid, a 
four-quark state, or a $K^* \bar{K}$ bound state 
\cite{e-kk}.  Because of its 
closeness in mass to the $\eta(1440)$ it is possible that the two 
states share a common origin.
\item[$f_J/\theta(1710)$] This state is seen in the gluon rich 
$J/\psi$ radiative decay in the $K\bar{K}$ final state but it is not 
seen in $K\bar{K}$ by the LASS collaboration.   
It is either spin 0 or spin 2.  
Its mass is not consistent with quark model expectations for a state 
with these quantum numbers and as a result is considered to be a prime 
glueball candidate \cite{f-glue} although, 
as already mentioned, another explanation 
is that it is a $VV$ molecule \cite{vv}.  
\item[$J^{PC}=0^{++}$] There are many more scalar mesons than can be 
accomodated as $q\bar{q}$ resonances.  For a long time the $a_0(980)$ 
and $f_0(980)$ were believed to be the isovector and $ns$ isovector 
members of the ground state scalar meson nonet.  However, the 
properties of these states were substantially different from the quark 
model predictions \cite{gi}
but were consistent with their interpretation as 
$K\bar{K}$ molecules analogous to the deuteron bound state \cite{iw}.  
With observation of the $a_0(1450)$ by the Crystal 
Barrel collaboration and the explanation of the $\pi\pi$ S-wave phase shift 
by the existence of (at least) the $f_0(980)$ and $f_0(1370)$ the 
$K\bar{K}$ molecule interpretation is reinforced.
\item[$f_0(1500)$] This state has been reported by the ASTERIX,
Crystal Barrel,
and GAMS Collaborations.  Because of its relatively narrow width,
its decay properties, and the fact its mass is consistent with lattice 
gauge theory glueball mass calculations \cite{f-glue,lat-glue}
the $f_0(1500)$ is considered to be a prime glueball candidate.
\item[$J^{PC}=2^{++}$]  The two $1^3P_2$ $q\bar{q}$ isoscalar states 
are most certainly the well known $f_2(1270)$ and $f_2'(1525)$.  
Numerous additional tensor mesons  have been observed which
have been suggested to be 
non-$q\bar{q}$ states: the $f_2(1430)$, $f_2(1520)$, $f_J(1710)$, 
$f_2(1810)$, etc \cite{pdb}.  
Clearly considerable work is needed to understand 
the nature of the mesons reported in this sector.
\end{description}

In addition, 
recent data gives hints of new states; the $a_1(1700)$ and the 
$\pi(1800)$.  There has been some speculation as to whether these 
are conventional $q\bar{q}$ resonances or  
hybrids \cite{barnes,close}.  
To distinguish the possibilities more data is needed.

To make progress in unravelling the meson spectrum by finding the 
missing states and solving some of these puzzles will take 
unprecedented statistics to perform the necessary 
partial wave analysis to filter 
the $J^{PC}$ quantum numbers.  In addition, techniques will have to be 
developed to study broad resonances, both theoretical and experimental.


\section{Meson Decays}

To test our understanding of mesons we have to go beyond a comparison 
of mass predictions and probe mesons' internal structure.  Because 
decays are sensitive to the details of the meson wavefunctions they 
are an important test of our understanding of the internal 
structure of mesons.  In addition, it is important to know the 
expected decay modes for meson searches.  As was already mentioned,  
knowing something about the expected decay properties might explain 
why they are missing.  Finally,  comparing the observed decay 
properties of mesons to the expectations of different interpretations, 
$g\bar{q}$ vs hybrid for example, is an important means of 
determining what they are.

\subsection{Decay Models}

A number of models exist in the literature; the pseudoscalar emmission 
model, the $^3P_0$ model, and the flux-tube breaking model \cite{i-k}.  
These decay
models all give good overall results but there can be large variations in 
individual decay rates depending on the specific assumptions of the 
calculation.  The results of a fit to well known decay widths are 
shown in Fig. 6.

\begin{figure}[h!]
\begin{center}
\begin{turn}{-90}
\epsfig{file=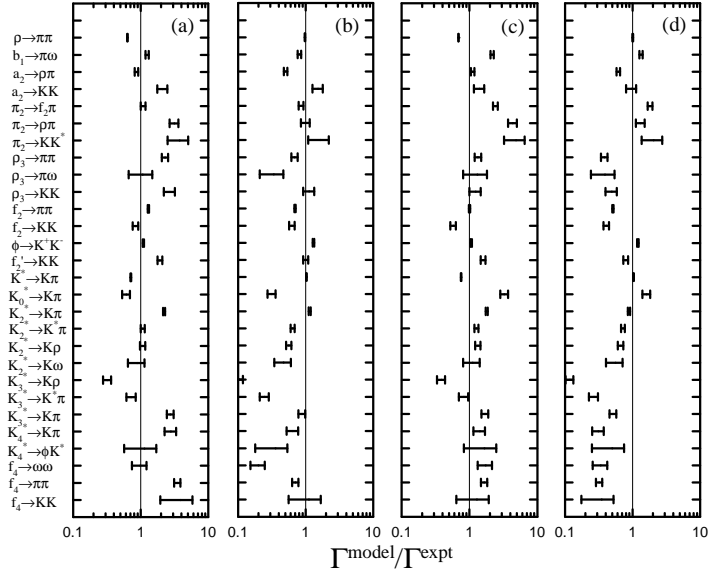,width=8.0cm,clip=}
\end{turn}
\end{center}
\caption{The ratio of decay model predictions for partial widths to 
the experimental values.  The error bars only include the effects of 
experimental errors in the ratios.  (a) and (b) correspond to 
variations of the  
$^3P_0$ model using SHO wave functions.  (c) and (d) correspond to 
variations of the flux-tube-breaking model using the wavefunctions
of Ref. \cite{gi}. }
\end{figure}

\subsection{The $\xi(2220)$: Strong Decays of the $1^3P_2$ and 
$1^3P_4$ $s\bar{s}$ Mesons}

The $\xi(2220)$ is an interesting example of the approach one takes 
to understand the nature of a new resonance.  The $\xi(2220)$ was 
discovered in $J/\psi$ radiative decay by the MARK III collaboration.  
It was found to be very narrow so on this basis it was assumed to be 
something exotic as conventional wisdom believed that a high mass 
conventional meson should be relatively broad.  Another interpretation 
pointed out that it could be an $s\bar{s}$ state as its masses were 
consistent with the masses predicted for the $1F$ $s\bar{s}$ mesons 
\cite{g-i-k}.  
A quark model calculation found the $^3F_J$ $s\bar{s}$ widths to be 
relatively narrow because these states have a limited number of decay 
modes which were calculated to be relatively narrow.  Unfortunately 
this study was not exhaustive and neglected decays to $L=1$ mesons 
under the assumption that with very little phase space these decay 
modes should be 
narrow.   The more important decays from a recent, more complete 
calculation, are given in Table 2 \cite{b-g}.

\begin{table}[h!]
\caption{Decay widths (in MeV) for the $^3F_4$ $s\bar{s}$ and
$^3F_2$ $s\bar{s}$ states. From Ref. \cite{b-g}.} 
\label{table3}
\begin{center}
\begin{tabular}{lc}
\hline
Decay  & $^3P_0$(KIPSN) \\
\hline
$f'_4 \to [K \bar{K}]_{0,G} $  & 29 \\
$f'_4 \to [K^*(892) \bar{K} +c.c.]_{1,G} $  & 27 \\
$f'_4 \to [K^*(892) \bar{K}^*(892)]_{2,D} $  & 44 \\
\hline
$\sum_i \Gamma_i$  & 132 \\
\hline \hline
$f'_2 \to [K \bar{K}]_{0,D} $  & 12 \\
$f'_2 \to [K^*(892) \bar{K} +c.c.]_{1,D} $  & 26 \\
{$f'_2 \to [K_1(1270) \bar{K}+c.c.]_{1,P} $}  & {187} \\
$f'_2 \to [K_1(1270) \bar{K}+c.c.]_{1,F} $  & 11 \\
{$f'_2 \to [K_2^*(1430) \bar{K}+c.c.]_{2,P} $}  & {24} \\
$f'_2 \to [K^*(892) \bar{K}^*(892)]_{2,G} $  & 12 \\
{$f'_2 \to [K_1(1270) \bar{K}^*(892) +c.c.]_{1,P} $}  & {40} \\
{$f'_2 \to [K_1(1270) \bar{K}^*(892) +c.c.]_{2,P} $} & {13}  \\
{$f'_2 \to [f_1(1510) \eta]_{1,P} $}  & {22} \\
\hline
$\sum_i \Gamma_i$ & 391  \\ \hline \hline
\end{tabular}
\end{center}
\end{table}

Including the previously neglected decay modes gives a much larger 
width for the $^3F_2$ state.  The $^3F_4$ width, although wider than 
the observed width, is still consistent within the large uncertainties 
of the model.  The dominant decay modes are to a pseudoscalar meson 
and a P-wave meson.  One can immediately see from Table 2 the reason 
for this.  When $^3F_4$ decays to S-wave mesons they are in a 
relatively high angular momentum state with the commensurate angular 
momentum barrier.  In contrast, when the $^3F_4$ decays to an S-wave 
and P-wave meson, the angular momentum can be obsorbed as  internal 
orbital angular momentum so that the mesons are emitted in a lower 
relative angular momentum state with the corresponding
smaller angular momentum barrier.  The  issues of
phase space is relevant comparing the width of the 
$K_1(1270)$ to the width to the $K_1(1400)$.  The lesson from this 
example is that it is extremely important to do a complete analysis 
which in this case means including all allowed decay modes in the 
calculation.

Recently the BES collaboration has reported measurements of this state 
with flavour symmetric decays \cite{bes}.  
If these results are confirmed they 
would reinforce the glueball explanation of this state.

\subsection{Additional Comments on Meson Decays}

We now have reasonably reliable tools for calculating decay widths which can 
be used to both understand the nature of new states and 
predict the most promising channels to search for the 
missing states \cite{i-k,barnes,close}.
In the former case the models are reasonably reliable, although not 
totally infallible, and could be used to distinguish between a 
conventional or hybrid description of a new resonance.  For example 
there are both conventional $q\bar{q}$ and hybrid 
predictions of an $\eta_2$ $J^{PC}=2^{-+}$ state around 1900~MeV.  
By comparing the predicted partial widths in both scenarios to 
observed widths one can distinguish between the two scenarios.

\begin{center}
\begin{tabular}{|l|r r r c r r r|}
\hline
	& $\rho\rho$ & $\omega\omega$ & $f_2 \eta$ & $a_0(1450)$ &
		$a_1 \pi$ & $a_2 \pi$ & $K^* K$ \\ \hline
$\eta_2(1D)$ & 147 & 46 & 45 & 1 & 43 & 264 & 61 \\
$\eta_2(H)$ & 0 & 0 & 20 & 2 & 0 & 160 & 10 \\
\hline
\end{tabular}
\end{center}

The interesting channels are the $\rho\rho$, $\omega\omega$, and $a_1 
\pi$ in which the $\eta_2(1D)$ has a large partial width while the 
$\eta_2(H)$ has a zero branching fraction.  Thus by looking in many 
different decay channels one can distinguish between the two 
possibilities.

For excited mesons we should expect cascade decays 
resulting in many particles in the final state. For example:
\begin{eqnarray}
1F & \to & 1D + \pi \\
& & \quad \searrow 1P + \pi \\
& & \quad \quad \searrow 1S + \pi
\end{eqnarray}
It will therefore be necessary to be able to reconstruct the initial 
resonance from the complicated many body final state.  An additional 
complication is that some of the intermediate states in the chain are likely 
to be broad.  It will be necessary to understand how to deal with this 
non trivial problem.

\section{Meson Photoproduction}

Before concluding I would like to make a few comments about meson 
production with intense high energy photons which
offer a number of photoproduction mechanisms to produce mesons.
\begin{description}
\item[Diffractive] 
This process is likely to be one of 
the more interesting production mechanisms.  It takes advantage of the 
vector meson content of the photon which can be described by vector 
meson dominance.  The vector meson can be excited into an excited 
state via pomeron exchange while the target remains in its ground state.  
What makes this process so interesting is 
that the photon has a relatively large $s\bar{s}$ content so that 
photoproduction offers the possibility of producing large numbers of 
excited $s\bar{s}$ mesons.  In some sense we can therefore regard the photon 
as a $\phi$ beam.  The $s\bar{s}$ mesons offer an 
intermediate step extrapolating from the $c\bar{c}$ states where we 
have some confidence in the quark model to light quark systems where 
the model is suspect.
\item[Inelastic]
This process is a variation of the previous process  except here the 
target is excited via, say, $\pi$ exchange.
\item[Charge Exchange]
This process is also similar except now the exchange particle is 
charged.
\item[Two Photon]
Here a photon is exchanged from the target and fuses with 
the incident photon to form the final state meson.  The two photon 
couplings offer one more piece of information that can be used to 
understand the nature of an observed state.  
\end{description}

A number of photoproduction experiments  have demonstrated the 
potential of this approach and have made intriguing observations;
SLAC hybrid facility, CERN $\Omega$ sepctrometer, FNAL 687, and HERA.
Clearly a dedicated facility with a high event rate has much to 
contribute to the subject.  

There are a number of issues that should be addressed.  A better 
theoretical understanding of photoproduction is necessary for 
planning experiments.  To this end, 
a useful starting point would be to survey  
the current state of theory.  This should be followed by new calculations.  
A second  topic needed to be dealt with as part of detector design is 
to consider what the signal will look like. The decay models find that
high L mesons will likely  cascade down to L-1 in steps leading to a 
large number of particles in the final state. Work should be started 
on theorist's simulations which can be fed into detector simulations.

\subsection{$\gamma\gamma \to M$}

Finally, I want to finish with some comments on two photon production 
of mesons.  
As was just stated, two-gamma couplings offer a useful probe of the 
internal structure of mesons \cite{barnes2}. 
The observation of a large $\gamma\gamma$ width for the 
$^1D_2$ $q\bar{q}$ state $\pi_2$ suggests that the $\gamma\gamma$ 
couplings of many orbitally excited light-quark $q\bar{q}$ states may 
be experimentally accessible.
 $\gamma\gamma$ couplings could also be 
crucial in establishing or refuting candidate gluonic mesons or other 
exotic candidates.  There are a number of non $q\bar{q}$ candidates 
that have been studied in this context;  $f_0(975)$, $a_0(980)$,
$\eta (1440)$, $f_2 (1720)$, and the $\xi(2220)$. 

The $\xi(2220)$ provides a timely example.
One typically expects $\Gamma_{\gamma\gamma}(M_{q\bar{q}}) \sim 
1$~keV.  ($\Gamma_{\gamma\gamma}(\pi_2 (1D)) \sim 0.1$~keV, 
($\Gamma_{\gamma\gamma}(\pi_4 (1G)) \sim 1$~keV).  
Recently the CLEO collaboration published a result placing the limit of 
$\Gamma_{\gamma\gamma} B_{KK} (\xi(2220)) \leq 1.3$~eV which gives 
$\Gamma_{\gamma\gamma} \leq 5$~eV \cite{cleo}.
This small two-photon width is much smaller than expected from quark 
model calculations and is another indication of the
non-$q\bar{q}$ nature of the $\xi$.

\section{Concluding Comments}

In this cursory overview of light $q\bar{q}$ resonances we have seen
that the constituent quark model generally gives a good description 
of light meson spectroscopy.  However, there are far too many missing 
states, especially radial and orbital excitations, and there are 
numerous puzzles that may point to physics beyond this simple picture 
of hadron spectroscopy.  Until we make progress on both fronts, by 
observing some of the missing states and making detailed measurements 
of the states in question, we cannot say that we fully understand 
hadron physics. To make progress in hadron model building we need
to find many states, both conventional and exotic to test and refine 
our models.
This will take unprecedented statistics to perform detailed partial 
wave analysis that can distinguish between the many different 
contributions to the cross section at high energy.
In addition theoretical 
progress is needed to understand exactly what is measured by 
experiment.  Some of the topics that need further study are coupled 
channel effects and final state interactions.  In addition it is time 
that the various production mechanisms be better understood.  
Effort needed towards the goal of building a 
dedicated meson spectrometer facility is a better understanding of
event rates  signals.

There is a compelling case for a dedicated meson spectroscopy 
facility.  The next step is to 
put the theoretical predictions in a form that is most useful to
experimentalists with the goal of designing a detector.

\section*{Acknowledgements}

The author thanks the organizers of the workshop for the invitation to 
participate and for providing a very stimulating environment.


\end{document}